\documentclass[11pt]{article}
\usepackage{amsmath}
\usepackage{amsfonts}
\usepackage{amssymb}
\usepackage{graphicx}
\usepackage{subfigure}

\def\bea{\begin{eqnarray}}
\def\eea{\end{eqnarray}}

\begin{document}
\begin{center}
\LARGE {\bf The effect of large-decoherence on mixing-time in
Continuous-time quantum walks on long-range interacting cycles
 }
\end{center}
\begin{center}
{\bf S. Salimi {\footnote {E-mail: shsalimi@uok.ac.ir}}, R. Radgohar {\footnote {E-mail: r.radgohar@uok.ac.ir}}}\\
 {\it Faculty of Science,  Department of Physics, University of Kurdistan, Pasdaran Ave., Sanandaj, Iran} \\
 \end{center}
\vskip 3cm
\begin{center}
{\bf{Abstract}}
\end{center}
In this paper, we consider decoherence in continuous-time quantum
walks on long-range interacting cycles (LRICs), which are the
extensions of the cycle graphs. For this purpose, we use Gurvitz's
model and assume that every node is monitored by the corresponding
point contact induced the decoherence process. Then, we focus on
large rates of decoherence and calculate the probability
distribution analytically and obtain the lower and upper bounds of
the mixing time. Our results prove that the mixing time is
proportional to the rate of decoherence and the inverse of the
distance parameter ($\emph{m}$) squared. This shows that the
mixing time decreases with increasing the range of interaction.
Also, what we obtain for $\emph{m}=0$ is in agreement with
Fedichkin, Solenov and Tamon's results~\cite{FST} for cycle, and
see that the mixing time of CTQWs on cycle improves with adding
interacting edges.

\newpage

\section{Introduction}
Random walks or Markov chains on graphs have broad applications in
various fields of mathematics~\cite{PD}, computer
sciences~\cite{MR} and the natural sciences, such as modeling of
crystals in solid state physics~\cite{JMZ}. The quantum walk(QW)
is a generalization of the classical random walk(CRW) developed by
exploiting the aspects of quantum mechanics, such as superposition
and interference~\cite{BN, AR}. The QW is widely studied in many
distinct fields, such as polymer physics, solid state physics,
biological physics and quantum computation~\cite{JK, GHW, BG}.
There are two distinct variants of QWs: the discrete-time
QWs(DTQWs)~\cite{AJVK, ADZ} and the continuous-time
QWs(CTQWs)~\cite{FG1}. In the CTQW, one can directly define the
walk on the position space, whereas in the DTQW, it is necessary
to introduce a quantum coin operation to define the direction in
which the particle has to move. Experimental implementations of
both QW variants have been reported e.g. on microwave
cavities~\cite{BCS}, ground state atoms~\cite{WD}, the orbital
angular momentum of photons~\cite{PZ} , waveguide
arrays~\cite{HBP} or Rydberg atoms~\cite{RCNJ, OMEA} in optical
lattices. Recently, quantum walks have been studied in numerous
publications. For instance, the DTQWs have been considered on
random environments~\cite{NKOD}, on quotient graphs~\cite{Krovi},
in phase space~\cite{PSMB} and for single and entangled
particles~\cite{CMCH}. Also, the CTQWs have been investigated on
the $\emph{n}$-cube~\cite{CMR}, on star graphs~\cite{ssa1, xp1},
on quotient graphs~\cite{SS}, on circulant Bunkbeds~\cite{LRS}, on
trees~\cite{KO}, on ultrametric spaces~\cite{KO2}, on odd
graphs~\cite{SSOG}, on simple one-dimensional lattices~\cite{ABT},
via modules of Bose-Mesner and Terwilliger algebras~\cite{SSM},
via spectral distribution associated with adjacency
matrix~\cite{JSMAA} and by using krylov subspace-Lanczos
algorithm~\cite{saj4}.

In this paper, we focus on the CTQWs on the one-dimensional
networks. These networks provide the better understanding of the
physical systems dynamics. For examples, they have been used to
describe the behavior of metals in solid state physics~\cite{JMZ},
to explain the dynamics of atoms in optical lattices~\cite{WD,
RCNJ}, to demonstrate Anderson localization in the systems with
energetic disorder~\cite{PWAP} and to address various aspects of
normal and anomalous diffusion~\cite{RMJK}. The simplest structure
of the one-dimensional networks generated by connecting the
nearest neighbors can be modeled the physical systems with
short-range interactions. Gravity and electromagnetism are the
only known fundamental interactions extending to a macroscopic
distance. Due to its basic importance, it has been along tradition
to search extra long-range interactions~\cite{CDHP}. In some of
the above mentioned experiments, e.g. the clouds of ultra-cold
Rydberg atoms assembled in a chain over which an exciton migrates,
the trapping of the exciton occurs at the ends of the chain, one
finds long-range interactions have to be considered~\cite{OMEA,
WARV}. Recently, it has been shown, from the view point of Quantum
Electrodynamics, the CTQW for all long-range interactions in a
linear system decaying as $R^{-\nu}$ (where $R$ is the distance
between two nodes of the network) belong to the same universality
class for $\nu>2$, while for classical continuous-time random walk
(CTRW) universality only holds for $\nu>3$~\cite{MPBE7}. It has
been shown the long-range interaction lead to a slowing-down of
the decay of the average survival probability, which is counter
intuitive since for the corresponding classical process one
observes a speed-up of the decay~\cite{MPBSE}. In~\cite{XP},
authors studied coherent exciton transport on a new network model,
namely long-range interacting cycles. LRICs are constructed by
connecting all the two nodes of distance $\emph{m}$ on the cycle
graphs. Since all the LRICs have the same value of connectivity
$k=4$ (the number of edges which exit from every node), thus LRICs
provide a good facility to study the influence of long range
interaction on the transport dynamics on various coupled dynamical
systems, including Josephson junction arrays~\cite{KWPB1},
synchronization~\cite{BBHPD}, small-world networks~\cite{DSNSW}
and many other self-organizing systems. All of the mentioned
articles have focused on a closed quantum system without any
interaction with its environment. Recently, more realistic
analysis of quantum walks by using decoherence concept has been
provided on line~\cite{RSA}, on circulant~\cite{KT, LT} and on
hypercube~\cite{WS}. In~\cite{FST}, the effect of decoherence in
the CTQWs on cycles has been considered analytically. The results
showed that, for small rates of decoherence, the mixing time
improves linearly with decoherence, whereas for large rates of
decoherence, the mixing time deteriorates linearly towards the
classical limit. In~\cite{SR}, the authors have studied the
influence of small decoherence in the CTQWs on long-range
interacting cycles(LRICs). The authors proved that the mixing time
is inversely proportional to the decoherence rate and also
independent of the distance parameter $\emph{m}$. Moreover, they
showed the mixing time upper bound for CTQWs on LRIC, remain close
to its value in the absence of shortcut links(cycle).  Now, we
want to investigate the effect of large decoherence on the mixing
time in the CTQWs on LRICs. For this end, we use of an analytical
model developed by Gurvitz~\cite{AG}. In this model, every vertex
is monitored by an individual point contact induced the
decoherence process.
 \clearpage
\begin{figure}[h]
\includegraphics{rr.eps}
\vspace{.035cm} \caption{Long-range interaction cycles $G(8,3)$ and
$G(10,4)$}\label{a}
\end{figure}
We calculate the probability distribution analytically then, for
large rates of decoherence, obtain the lower and upper bounds of the
mixing time. Our analytical results prove the bounds of the mixing
time are proportional to the decoherence rate that in agreement with
~\cite{FST}'s results . Moreover, we prove that these bounds are
inversely proportional to square of the distance parameter
$\emph{m}$, i.e. the mixing time decreases with increasing $m$.
\\This paper is organized as follows: In Sec. 2, we briefly describe
the network structure LRIC. The CTQWs on LRICs are considered in
Sec. 3. In Sec. 4, we study the effect of decoherence in CTQWs on LRICs,
then we focus on the large rates of decoherence and calculate the
probability distribution in Sec. 5. In Sec. 6, we define the mixing time
and obtain its lower and upper bounds. We conclude with a summery
of results in Sec. 7.

\section{Structure of LRIC}
Long-range interacting cycle (LRIC) can be generated as
follows~\cite{XP}: First, the network be composed of a cycle graph.
Second, two nodes of distance $\emph{m}$ on the cycle graph are
linked by an additional bond. We continue the second step until all
the two nodes of distance $\emph{m}$ have been connected. LRIC
denoted by $G(N, \emph{m})$, is characterized by the network size
$N$ and the long-range interaction parameter $\emph{m}$. Fig. 1
shows the sketches of $G(8, 3)$ and $G(10, 4)$.

\section{CTQWs on LRICs}
In general, every network is characterized by a graph whose bonds
connect nodes with a wide distribution of mutual distances.
Algebraically, every graph corresponds with a discrete Laplace
operator $\textbf{A}$. We introduce the states $|j\rangle$ which are
localized at the nodes $j$ of the graph and take the set
$\{|j\rangle\}$ to be orthonormal. We assume that transition rates
$\gamma$ between all nodes are equal and $\gamma\equiv 1$. The
non-diagonal element of matrix $\textbf{A}$ ($A_{ij}$) equals 1 if
nodes $i$ and $j$ are connected by a bond and 0 otherwise. The
diagonal element $\textbf{A}_{ii}$ equals $-k_{i}$ that $k_{i}$ is
the number of bonds which exit from node $i$\textbf{~\cite{FG1,
CHGO}}. Since the states $|j\rangle$ span the whole Hilbert space,
the time evolution of a state $|j\rangle$ starting at time 0 is
determined by the system Hamiltonian $H=A$ as
$|j,t\rangle=U(t)|j\rangle$, where $U(t)=\exp[-iHt]$ is the quantum
mechanical time evolution operator\textbf{~\cite{FG1, CHGO}}. Thus
the Hamiltonian matrix $H$ of $G(N,\emph{m})$ $(m\geq2)$ can be
written as

\begin{eqnarray}\label{1}
   H_{ij}=\langle i|H|j \rangle=\left\{
                                \begin{array}{ll}
                                  -4, & \hbox{if $i=j$;} \\
                                  1, & \hbox{if $i=j\pm 1$;} \\
                                  1, & \hbox{if $i=j\pm m$;} \\
                                  0, & \hbox{Otherwise.}
                                \end{array}
                              \right.
        \end{eqnarray}

The Hamiltonian acting on the state $|j\rangle$ is given by
\begin{eqnarray}\label{2}
   H|j\rangle=-4|j\rangle+|j-1\rangle+|j+1\rangle+|j-m\rangle+|j+m\rangle.
   \end{eqnarray}
The above equation is the discrete version of the Hamiltonian for a
free particle moving on the cycle. Using Bloch function~\cite{JMZ,
CK} in solid state physics, the time independent Schr\"{o}dinger
equation can be written as
\begin{eqnarray}\label{3}
   H|\psi_{n}\rangle=E_{n}|\psi_{n}\rangle.
   \end{eqnarray}
The Bloch states $|\psi_{n}\rangle$ can be expanded as a linear
combination of states $|j\rangle$
\begin{eqnarray}\label{4}
   |\psi_{n}\rangle=\frac{1}{\sqrt{N}}\sum_{j=1}^{N}e^{-i\theta_{n} j}|j\rangle.
   \end{eqnarray}
Substituting Eqs. (2) and (4) into Eq. (3), we obtain the
eigenvalues of the system as
\begin{eqnarray}\label{5}
  E_{n}= -4+2\cos(\theta_{n})+2\cos(m\theta_{n}).
   \end{eqnarray}
\\The Bloch relation can be obtained by
projecting $|\psi_{n}\rangle$ on the state $|j\rangle$ such that
$\psi_{n}(j)\equiv\langle j|\psi_{n}\rangle=e^{-i(\theta_{n} j)}/
N$. It follows that $\theta_{n}=2n\pi/ N$ with $n$ integer and
$n\in[0,N)$. From Schr\"{o}dinger equation, we have
\begin{eqnarray}\label{6}
 i\hbar\frac{d}{dt}|\psi_{n}(t)\rangle=H|\psi_{n}(t)\rangle.
   \end{eqnarray}
By assuming $\hbar=1$ and $|\psi_{n}(0)\rangle=|0\rangle$, the
solution of the above equation is
$|\psi_{n}(t)\rangle=e^{-iHt}|0\rangle$.
\\The probability to find the walker at node $j$ at time $t$
 is given by \\$P_{j}(t)=|\langle j|\psi_{n}(t)\rangle|^{2}$.

\section{The Decoherent CTQWs on LRICs}
Here, we want to study the appearance of decoherence in the CTQWs
and obtain the probability distribution $P(t)$. For this end, we
make use of an analytical model developed by Gurvitz~\cite{AG, GFM}.
In this model, a ballistic point-contact is placed near each node of
network that is taken as noninvasive detector. Gurvitz demonstrated
that measurement process is fully described by the Bloch-type
equations applied to whole system. These equations led to the
collapse of the density-matrix into the statistical mixture in the
course of the measurement process.
\\According to Gurvitz model, the time evolution of
density matrix for our network (LRIC), has the following form
 \begin{eqnarray}\label{7}
\begin{array}{cc}
  \frac{d}{dt}\rho_{j,k}(t)= & \frac{i}{4}[-\rho_{j-1,k}-\rho_{j+1,k}-\rho_{j-m,k}-\rho_{j+m,k}+\rho_{j,k-1}  \\
   &  \\
   &
   +\rho_{j,k+1}+\rho_{j,k-m}+\rho_{j,k+m}]-\Gamma(1-\delta_{j,k})\rho_{j,k},
\end{array}
  \end{eqnarray}

that density matrix $\rho(t)$ is as
$\rho(t)=|\psi(t)\rangle\langle\psi(t)|$ and $\Gamma$ is the
decoherence parameter. Also, we observe that
$P_{j}(t)=\rho_{j,j}(t)$.
\\We define the variable $S_{j,k}$ as ~\cite{FST}
\begin{eqnarray}\label{8}
 S_{j,k}=i^{k-j}\rho_{j,k}
   \end{eqnarray}
and by substituting it into Eq. (7), we obtain
\begin{eqnarray}\label{9}
\begin{array}{cc}
  \frac{d}{dt}S_{j,k}=  & \frac{1}{4}[-S_{j-1,k}+S_{j+1,k}-i^{-m+1}S_{j-m,k}-i^{m+1}S_{j+m,k}-S_{j,k-1} \\
   &\\
 &
 +S_{j,k+1}+i^{m+1}S_{j,k-m}+i^{-m+1}S_{j,k+m}]-\Gamma(1-\delta_{j,k})S_{j,k}.
\end{array}
    \end{eqnarray}
Note that by assuming $\emph{m}=0$, we achieve the relations
mentioned in~\cite{FST}.

\section{Large Decoherence}
Firstly, we review the results obtained in~\cite{SR} for the
decoherent CTQWs on LRICs with small rate of decoherence. The
mixing time upper bound for the odd values of $m$ is as
$T_{mix}(\epsilon)<\frac{1}{\Gamma}\ln(\frac{N}{\epsilon})[\frac{N}{N-2}]$
and for the even $m$s is as
$T_{mix}(\epsilon)\leq\frac{1}{\Gamma}\ln(\frac{N}{\epsilon
})[\frac{N}{N-1}]$. Thus, the mixing time upper bound for odd $m$
is larger than the mixing time upper bound for even $m$. In
addition to these relations show the upper bound of the mixing
time is inversely proportional to decoherence rate $\Gamma$ and
independent of the distance parameter $m$.
\\In the following, we assume that the rate
decoherence $\Gamma$ is large ($\Gamma\gg 1$) and consider its
effect in CTQWs on LRICs.
\\Firstly, as mentioned in~\cite{FST}, we define the diagonal sum
$D_{k}$ as
\begin{eqnarray}\label{10}
 D_{k}=\sum_{j=0}^{N-1}S_{j,j+k(mod N)},
   \end{eqnarray}
where the indices are treated as integers modulo $N$. \\From Eq.
(9), one can achieve the following form
\begin{eqnarray}\label{11}
 \frac{d}{dt}D_{k}=-\Gamma(1-\delta_{k,0})D_{k}.
   \end{eqnarray}
Thus the minor diagonal sums ($D_{k}$ for $k\neq 0$) decay with
characteristic time of order $1/\Gamma$. \\Also, by applying Eq. (9)
for the elements on the two minor diagonals nearest to the major
diagonal, we observe that these elements make limit to small values
of order $1/\Gamma$.
\\Using the above way for the other diagonals, it is yielded the
secondary set of matrix elements in terms of nearness to major
diagonal will be of the order of $1/\Gamma^{2}$, etc.
\\Now, we consider only matrix elements of order of
$1/\Gamma$ and get

\begin{eqnarray}\label{12}
\begin{array}{cc}
 S^{\prime}_{j,j}= & \frac{1}{4}[-S_{j-1,j}+S_{j+1,j}-i^{-m+1}S_{j-m,j}-i^{m+1}S_{j+m,j} \\
  & \\
   & -S_{j,j-1}+S_{j,j+1}+i^{m+1}S_{j,j-m}+i^{-m+1}S_{j,j+m}],
\end{array}
   \end{eqnarray}

\begin{eqnarray}\label{13}
 S^{\prime}_{j,j+1}=\frac{1}{4}[S_{j+1,j+1}-S_{j,j}]-\Gamma S_{j,j+1},
   \end{eqnarray}

\begin{eqnarray}\label{14}
 S^{\prime}_{j,j-1}=\frac{1}{4}[-S_{j-1,j-1}+S_{j,j}]-\Gamma S_{j,j-1},
   \end{eqnarray}

\begin{eqnarray}\label{15}
 S^{\prime}_{j,j+m}=\frac{1}{4}[-i^{m+1}S_{j+m,j+m}+i^{m+1}S_{j,j}]-\Gamma
 S_{j,j+m}
   \end{eqnarray}

\begin{eqnarray}\label{16}
 S^{\prime}_{j,j-m}=\frac{1}{4}[-i^{-m+1}S_{j-m,j-m}+i^{-m+1}S_{j,j}]-\Gamma
 S_{j,j-m}.
   \end{eqnarray}

For simplicity's sake, we define
\begin{eqnarray}\label{17}
\begin{array}{ccc}
  a_{j}=S_{j,j}, & d_{j}=S_{j,j+1}+S_{j+1,j}, &
  f_{j}=i^{-m+1}S_{j,j+m}-i^{m+1}S_{j+m,j}.
\end{array}
   \end{eqnarray}

After some algebra, one can get

\begin{eqnarray}\label{18}
 a^{\prime}_{j}=\frac{1}{4}[-d_{j-1}+d_{j}+f_{j}-f_{j-m}],
   \end{eqnarray}

\begin{eqnarray}\label{19}
 d^{\prime}_{j}=\frac{1}{2}[a_{j+1}-a_{j}]-\Gamma d_{j},
   \end{eqnarray}

\begin{eqnarray}\label{20}
 f^{\prime}_{j}=\frac{1}{2}[a_{j+m}-a_{j}]-\Gamma f_{j}.
   \end{eqnarray}
Differentiation of the above equation gives

\begin{eqnarray}\label{21}
 a^{\prime \prime}_{j}=\frac{1}{4}[-d_{j-1}^{\prime}+d_{j}^{\prime}+f_{j}^{\prime}-f_{j-m}^{\prime}],
   \end{eqnarray}

\begin{eqnarray}\label{22}
 d^{\prime \prime}_{j}=\frac{1}{2}[a_{j+1}^{\prime}-a_{j}^{\prime}]-\Gamma d_{j}^{\prime},
   \end{eqnarray}

\begin{eqnarray}\label{23}
 f^{\prime \prime}_{j}=\frac{1}{2}[a_{j+m}^{\prime}-a_{j}^{\prime}]-\Gamma f_{j}^{\prime}.
   \end{eqnarray}

 Let us now assume that we have the solutions of Eqs.
(21), (22) and (23) as

\begin{eqnarray}\label{24}
\begin{array}{ccc}
  a_{j}=\displaystyle\sum_{k=0}^{N-1}A_{k}e^{\frac{2\pi jk}{N}}e^{-\gamma_{k} t}, & d_{j}=\displaystyle\sum_{k=0}^{N-1}D_{k}e^{\frac{2\pi jk}{N}}
  e^{-\gamma_{k} t}, & f_{j}=\displaystyle\sum_{k=0}^{N-1}F_{k}e^{\frac{2\pi jk}{N}}e^{-\gamma_{k}
  t},
\end{array}
   \end{eqnarray}

that $\gamma_{k}$, $A_{k}$, $D_{k}$ and $F_{k}$ are unknown.
\\To obtain $\gamma_{k}$, we substitute these solutions into the above mentioned
equations.
\begin{eqnarray}\label{25}
\begin{array}{c}
  A\gamma_{k}^{2}+D(\frac{\gamma_{k}}{4}(-e^{\frac{-2\pi ik}{N}}+1))+F(\frac{\gamma_{k}}{4}(-e^{\frac{-2\pi ikm}{N}}+1))=0, \\
   \\
  A(\frac{\gamma_{k}}{2}(e^{\frac{2\pi ik}{N}}-1))+D(\gamma_{k}^{2}-\gamma_{k}\Gamma)=0,\\
  \\
  A(\frac{\gamma_{k}}{2}(e^{\frac{2\pi ikm}{N}}-1))+F(\gamma_{k}^{2}-\gamma_{k}\Gamma)=0.
\end{array}
   \end{eqnarray}

Note that for having nontrivial solutions, the determinant
of the coefficients matrix must be zero. Thus, we get
\begin{eqnarray}\label{25}
 \gamma_{k}^{2}(\gamma_{k}^{2}-\gamma_{k} \Gamma)[\gamma_{k}^{2}-\gamma_{k}
\Gamma+\frac{1}{2}(\sin^{2}(\frac{\pi k}{N})+\sin^{2}(\frac{\pi
km}{N}))]=0,
   \end{eqnarray}

and achieve
\begin{eqnarray}\label{26}
 \gamma_{k}=\left\{
          \begin{array}{ll}
            \gamma_{k,0}=\frac{1}{2\Gamma}(\sin^{2}(\frac{\pi k}{N})+\sin^{2}(\frac{\pi km}{N})), & \hbox{} \\
            \gamma_{k,1}=\Gamma -\frac{1}{2\Gamma}(\sin^{2}(\frac{\pi k}{N})+\sin^{2}(\frac{\pi km}{N})), & \hbox{} \\
            \gamma_{k,2}=0, & \hbox{} \\
            \gamma_{k,3}=\Gamma. & \hbox{}
          \end{array}
        \right.
 \end{eqnarray}

The general solutions of Eq. (21), (22) and (23) are as
\begin{eqnarray}\label{27}
 a_{j}=\frac{1}{N}\sum_{k=0}^{N-1}\{A_{k,0}e^{-\gamma_{k,0}t}+A_{k,1}e^{-\gamma_{k,1}t}+A_{k,2}e^{-\gamma_{k,2}t}+A_{k,3}e^{-\gamma_{k,3}t}\}\omega^{jk},
   \end{eqnarray}

\begin{eqnarray}\label{28}
d_{j}=\frac{1}{N}\sum_{k=0}^{N-1}\{D_{k,0}e^{-\gamma_{k,0}t}+D_{k,1}e^{-\gamma_{k,1}t}+D_{k,2}e^{-\gamma_{k,2}t}+D_{k,3}e^{-\gamma_{k,3}t}\}\omega^{jk},
   \end{eqnarray}

\begin{eqnarray}\label{29}
f_{j}=\frac{1}{N}\sum_{k=0}^{N-1}\{F_{k,0}e^{-\gamma_{k,0}t}+F_{k,1}e^{-\gamma_{k,1}t}+F_{k,2}e^{-\gamma_{k,2}t}+F_{k,3}e^{-\gamma_{k,3}t}\}\omega^{jk},
   \end{eqnarray}
where $\omega=e^{\frac{2\pi i}{N}}$. We replace $\gamma_{k}$s in
Eqs. (25) and use of the initial conditions $a_{j}=\delta_{j,0}$ and
$d_{j}=0$ for $j=0,...,N-1$. We set the constant coefficients into
the others and obtain
\begin{eqnarray}\label{30}
\begin{array}{ccc}
  A_{k,0}\simeq1, & A_{k,1}\simeq\frac{-1}{2\Gamma^{2}}(\sin^{2}(\frac{\pi k}{N})+\sin^{2}(\frac{\pi
km}{N})), & A_{k,3}=0
,
\end{array}
   \end{eqnarray}

\begin{eqnarray}\label{31}
\begin{array}{cc}
  D_{k,0}\simeq\frac{i}{\Gamma}\sin(\frac{\pi k}{N})\exp(\frac{i\pi
k}{N}),  &
D_{k,1}\simeq\frac{-i}{\Gamma}\sin(\frac{\pi k}{N})\exp(\frac{i\pi
k}{N}),\\
   &
\end{array}
\\D_{k,3}\simeq\frac{i}{\Gamma^{2}}\sin(\frac{\pi
k}{N})\exp(\frac{i\pi k}{N}),\hspace{2cm}\nonumber
   \end{eqnarray}

\begin{eqnarray}\label{32}
\begin{array}{cc}
  F_{k,0}\simeq\frac{i}{\Gamma}\sin(\frac{\pi km}{N})\exp(\frac{i\pi km}{N}),  & F_{k,1}\simeq\frac{-i}{\Gamma}\sin(\frac{\pi km}{N})\exp(\frac{i\pi
km}{N}),\\
   &
\end{array}
\\F_{k,3}\simeq\frac{i}{\Gamma^{2}}\sin(\frac{\pi km}{N})\exp(\frac{i\pi km}{N}).\hspace{2cm}\nonumber
   \end{eqnarray}

The full solution for $S(t)$ has the form
as
\begin{eqnarray}\label{33}
 S_{j,k}(t)=\left\{
              \begin{array}{ll}
                a_{j}, & \hbox{if j=k;} \\
                d_{j}/2, & \hbox{if  $j=k\pm 1$;} \\
                f_{j}/2, & \hbox{if  $j=k\pm m$;} \\
                0, & \hbox{otherwise.}
              \end{array}
            \right.
   \end{eqnarray}

Thus, the probability distribution $P_{j}(t)$ is the same $a_{j}$.
At large rates of decoherence $\Gamma$, Eq. (28) reduce to
\begin{eqnarray}\label{34}
 a_{j}(t)=\frac{1}{N}\sum_{k=0}^{N-1}\{\exp[\frac{-1}{2\Gamma}(\sin^{2}(\frac{\pi
k}{N})+\sin^{2}(\frac{\pi km}{N}))t]\}\omega^{jk}
   \end{eqnarray}

One can see the above distribution do not converge to any
stationary distribution. The reason being that the evolution of
the QW, as mentioned in Sec. 3, is given by the unitary operator
$U=e^{-itH}$. This is the fact that unitary operators preserve the
norm of states, and hence the distance between the states
describing the system at subsequent times does not converge to
zero[9]. This implies that the probability distribution of CTQW
does not converge to CRW.
\section{The bounds of mixing time}
Mixing time is the time it takes for the walk to approximate the
uniform distribution~\cite{FST, AJVK}, i.e.
\begin{eqnarray}\label{35}
T_{mix}=min\{T :
\sum_{j=0}^{N-1}|P_{j}(t)-\frac{1}{N}|\leq\epsilon\}.
   \end{eqnarray}
From Eq. (35), we obtain
\begin{eqnarray}\label{36}
\sum_{j=0}^{N-1}|a_{j}(t)-\frac{1}{N}|=
 \sum_{j=0}^{N-1}|\frac{1}{N}\sum_{k=0}^{N-1}\exp[\frac{-1}{2\Gamma}(\sin^{2}(\frac{\pi
k}{N})+\sin^{2}(\frac{\pi km}{N}))t]e^{\frac{2\pi
ijk}{N}}-\frac{1}{N}|,
   \end{eqnarray}
that simplifies to
\begin{eqnarray}\label{37}
\sum_{j=0}^{N-1}|
a_{j}(t)-\frac{1}{N}|=\frac{1}{N}\sum_{j=0}^{N-1}|\sum_{k=1}^{N-1}\exp[\frac{-1}{2\Gamma}(\sin^{2}(\frac{\pi
k}{N})+\sin^{2}(\frac{\pi km}{N}))t]\cos(\frac{2\pi kj}{N})|.
   \end{eqnarray}

\textbf{Lower bound}
\\To obtain a lower bound, we apply the following way:
\\We retain only the terms $j=0$ and $k=1, N-1$.
\begin{eqnarray}\label{38}
\begin{array}{cc}
  \displaystyle\sum_{j=0}^{N-1}|a_{j}(t)-\frac{1}{N}|\geq & |a_{0}(t)-\frac{1}{N}|=\frac{1}{N}\displaystyle\sum_{k=1}^{N-1}\exp(\frac{-(\sin^{2}(\frac{\pi
k}{N})+\sin^{2}(\frac{\pi km}{N}))t}{2\Gamma}) \\
   &\\
   \geq\hspace{-2.80cm}& \displaystyle\frac{2}{N}\exp(\frac{-(\sin^{2}(\frac{\pi }{N})+\sin^{2}(\frac{\pi
m}{N}))t}{2\Gamma}).\hspace{3cm}
\end{array}
   \end{eqnarray}

According to the mixing time definition, we have
\begin{eqnarray}\label{39}
T_{lower}=\frac{2\Gamma}{\sin^{2}(\frac{\pi }{N})+\sin^{2}(\frac{\pi
m}{N})}\ln(\frac{2}{N\epsilon}),
   \end{eqnarray}

and for $N\gg1$
\begin{eqnarray}\label{40}
T_{lower}\simeq\frac{2\Gamma
N^{2}}{\pi^{2}(1+m^{2})}\ln(\frac{2}{N\epsilon}).
   \end{eqnarray}

Note that for $m=0$,
\begin{eqnarray}\label{41}
T_{lower}\simeq\frac{2\Gamma
N^{2}}{\pi^{2}}\ln(\frac{2}{N\epsilon}),
   \end{eqnarray}
that is the same~\cite{FST}'s result. One observe the mixing time
lower bound in CTQWs on cycle decreases with adding interacting
links.

 \textbf{Upper bound}
\\Now, we want to obtain upper bound as following
\begin{eqnarray}\label{42}
\begin{array}{cc}
  \displaystyle\sum_{j=0}^{N-1}|a_{j}(t)-\frac{1}{N}|&= \frac{1}{N}\displaystyle\sum_{j=0}^{N-1}\mid\displaystyle\sum_{k=1}^{N-1}\exp[\frac{-1}{2\Gamma}(\sin^{2}(\frac{\pi
k}{N})+\sin^{2}(\frac{\pi km}{N}))t]\cos(\frac{2\pi kj}{N})\mid\\
   &  \\
    \leq\hspace{-3.5cm}&\frac{1}{N}\displaystyle\sum_{j=0}^{N-1}\displaystyle\sum_{k=1}^{N-1}\exp(\frac{-(\sin^{2}(\frac{\pi
k}{N})+\sin^{2}(\frac{\pi km}{N}))}{2\Gamma})\hspace{2.60cm}
\end{array}
   \end{eqnarray}
\\
Since for $0< x <\frac{\pi}{2}$, there is $\sin x>\frac{2x}{\pi}$~\cite{MASH}, we
get
\begin{eqnarray}\label{43}
\begin{array}{cc}
  \displaystyle\sum_{j=0}^{N-1}|a_{j}(t)-\frac{1}{N}|\leq & \frac{2}{N}\displaystyle\sum_{j=0}^{N-1}\displaystyle\sum_{k=1}^{[N/2]}\exp(\frac{-1}{\Gamma}(\sin^{2}(\frac{\pi
k}{N})+\sin^{2}(\frac{\pi km}{N}))t)\hspace{1cm} \\
   &  \\
   \leq\hspace{-3cm}& \frac{2}{N}\displaystyle\sum_{j=0}^{N-1}\displaystyle\sum_{k=1}^{[N/2]}\exp(\frac{-1}{\Gamma}(\frac{2k^{2}+2k^{2}m^{2}}{N^{2}})t)\hspace{2.6cm} \\
   &  \\
   \leq\hspace{-3cm}& \frac{2}{N}\displaystyle\sum_{j=0}^{N-1}\displaystyle\sum_{k=1}^{[N/2]}\exp(\frac{-1}{\Gamma}(\frac{2k+2km}{N^{2}})t)\hspace{3cm} \\
   &  \\
   \leq\hspace{-3cm}&
\frac{2}{N}\displaystyle\sum_{j=0}^{N-1}\displaystyle\sum_{k=1}^{\infty}\exp(\frac{-1}{\Gamma}(\frac{2k+2km}{N^{2}})t)\hspace{3cm}
\end{array}
   \end{eqnarray}
that in the third inequality, we used of the relation $k^{2}\geq k$
for $k\geq1$.
\begin{eqnarray}\label{44}
\sum_{j=0}^{N-1}|a_{j}(t)-\frac{1}{N}|\leq\frac{2}{\exp[{\frac{-1}{\Gamma}(\frac{2}{N^{2}}+\frac{2m^{2}}{N^{2}}})]-1}
   \end{eqnarray}
Thus,
\begin{eqnarray}\label{45}
T_{upper}=\frac{\Gamma
N^{2}}{2(1+m^{2})}\ln(\frac{2+\epsilon}{\epsilon}).
   \end{eqnarray}
Note that for $m=0$, we get
\begin{eqnarray}\label{46}
T_{upper}=\frac{\Gamma N^{2}}{2}\ln(\frac{2+\epsilon}{\epsilon}).
   \end{eqnarray}
   which is in agreement with~\cite{FST}'s result.
One can see that the mixing time upper bound for LRIC is smaller
than the mixing time upper bound for cycle. Based on the above
analysis, the mixing time in CTQWs on cycle decreases with adding
interacting links. Moreover, we observe that the mixing time is
proportional to the square of distance parameter $\emph{m}$ such
that it decreases with increasing the interaction range. Now, we
want to compare the lower bound with the upper bound. Since Eq.
(41) was obtained by assuming $N\gg1$ and $\epsilon\ll 1$, we have
$\ln(\frac{2}{N\epsilon})<\ln(\frac{2+\epsilon}{\epsilon}$). Thus,
we can write
\begin{eqnarray}\label{47}
\frac{2\Gamma
N^{2}}{\pi^{2}(1+m^{2})}\ln(\frac{2}{N\epsilon})<T_{mix}<\frac{\Gamma
N^{2}}{2(1+m^{2})}\ln(\frac{2+\epsilon}{\epsilon}).
   \end{eqnarray}
\section{Conclusion}
We studied the continuous-time quantum walk on long-range
interacting cycles (LRICs) under large decoherence $\Gamma\gg 1$.
We obtained the probability distribution analytically and found
the mixing time is bounded as
\begin{eqnarray}\label{47}
\frac{2\Gamma
N^{2}}{\pi^{2}(1+m^{2})}\ln(\frac{2}{N\epsilon})<T_{mix}<\frac{\Gamma
N^{2}}{2(1+m^{2})}\ln(\frac{2+\epsilon}{\epsilon}).
\end{eqnarray}

We proved that the $T_{mix}$ is proportional to decoherence rate
$\Gamma$ that this result accord with the conclusion is obtained
in~\cite{FST}. In~\cite{SR} has been proved that $T_{mix}$, for
small rates of decoherence, is independent of distance parameter
$\emph{m}$, while in this paper we showed that $T_{mix}$, for
large rates of decoherence, is inversely proportional to square of
$\emph{m}$. In other words, the mixing time decreases with
increasing the range of interaction. Also, since LRIC is the same
generalized cycle, by replacing $\emph{m}=0$ in all the above
relations, one achieves~\cite{FST}'s results. Moreover, we proved
the mixing times improve with adding interacting links.

\end{document}